\pdfoutput=1

\documentclass[preprint]{vgtc}               

\graphicspath{{./}} 

\usepackage{times}                     

\usepackage{mathptmx}                  

\vgtcinsertpkg

\ieeedoi{10.1109/VIS55277.2024.00052}

\usepackage{csquotes}
\usepackage{tikz}
\usetikzlibrary{calc}
\usepackage{xfrac}
\usepackage{subcaption}
\usepackage{tabularray}

\definecolor{fhg}{HTML}{179C7D}
\definecolor{fhglite}{HTML}{B4DDD3}

\title{Towards a Quality Approach to Hierarchical Color Maps}

\author{Tobias Mertz\thanks{e-mail: tobias.mertz@igd.fraunhofer.de}\\ %
        \scriptsize Fraunhofer IGD %
\and Jörn Kohlhammer\thanks{e-mail: joern.kohlhammer@igd.fraunhofer.de}\\ %
     \parbox{1.4in}{\scriptsize \centering Fraunhofer IGD \\ TU Darmstadt}}

\teaser{
  \centering
\normalfont
\begin{subfigure}{0.19\textwidth}
    \centering
    \begin{tikzpicture}[x=0.8em,y=0.8em]
    {
        \definecolor{n0}{HTML}{ee944c}
        \definecolor{n00}{HTML}{ef5f5b}
        \definecolor{n01}{HTML}{d37829}
        \definecolor{n02}{HTML}{a68f00}
        \definecolor{n020}{HTML}{8a7500}
        \definecolor{n0200}{HTML}{795900}
        \definecolor{n0201}{HTML}{656100}
        \definecolor{n1}{HTML}{00c6a9}
        \definecolor{n10}{HTML}{00a961}
        \definecolor{n100}{HTML}{008f45}
        \definecolor{n11}{HTML}{00abbb}
        \definecolor{n110}{HTML}{0091a3}
        \definecolor{n1100}{HTML}{00777a}
        \definecolor{n1101}{HTML}{00769d}
        \definecolor{n2}{HTML}{ba9aff}
        
        \filldraw[fill=n0] (0:1) -- (0:2) arc(0:120:2) -- (120:1) arc(120:0:1);
        \filldraw[fill=n00] (0:2) -- (0:3) arc(0:40:3) -- (40:2) arc(40:0:2);
        \filldraw[fill=n01] (40:2) -- (40:3) arc(40:80:3) -- (80:2) arc(80:40:2);
        \filldraw[fill=n02] (80:2) -- (80:3) arc(80:120:3) -- (120:2) arc(120:80:2);
        \filldraw[fill=n020] (80:3) -- (80:4) arc(80:120:4) -- (120:3) arc(120:80:3);
        \filldraw[fill=n0200] (80:4) -- (80:5) arc(80:100:5) -- (100:4) arc(100:80:4);
        \filldraw[fill=n0201] (100:4) -- (100:5) arc(100:120:5) -- (120:4) arc(120:100:4);
        \filldraw[fill=n1] (120:1) -- (120:2) arc(120:240:2) -- (240:1) arc(240:120:1);
        \filldraw[fill=n10] (120:2) -- (120:3) arc(120:180:3) -- (180:2) arc(180:120:2);
        \filldraw[fill=n100] (120:3) -- (120:4) arc(120:180:4) -- (180:3) arc(180:120:3);
        \filldraw[fill=n11] (180:2) -- (180:3) arc(180:240:3) -- (240:2) arc(240:180:2);
        \filldraw[fill=n110] (180:3) -- (180:4) arc(180:240:4) -- (240:3) arc(240:180:3);
        \filldraw[fill=n1100] (180:4) -- (180:5) arc(180:210:5) -- (210:4) arc(210:180:4);
        \filldraw[fill=n1101] (210:4) -- (210:5) arc(210:240:5) -- (240:4) arc(240:210:4);
        \filldraw[fill=n2] (240:1) -- (240:2) arc(240:360:2) -- (360:1) arc(360:240:1);

        \node at (-3.5em,3.5em) {\textbf{(a)}};
    }
    \end{tikzpicture}
\end{subfigure}
\begin{subfigure}{0.19\textwidth}
    \centering
    \begin{tikzpicture}[x=0.8em,y=0.8em]
        {
        \definecolor{n0}{HTML}{d0b099}
        \definecolor{n00}{HTML}{d69790}
        \definecolor{n01}{HTML}{ca9d7d}
        \definecolor{n02}{HTML}{b5a675}
        \definecolor{n020}{HTML}{a89756}
        \definecolor{n0200}{HTML}{a18537}
        \definecolor{n0201}{HTML}{938a36}
        \definecolor{n1}{HTML}{8fc1b6}
        \definecolor{n10}{HTML}{79b493}
        \definecolor{n100}{HTML}{57a87d}
        \definecolor{n11}{HTML}{62b5b9}
        \definecolor{n110}{HTML}{23a9af}
        \definecolor{n1100}{HTML}{009d9a}
        \definecolor{n1101}{HTML}{009cb0}
        \definecolor{n2}{HTML}{bdb2d5}
        
        \filldraw[fill=n0] (0:1) -- (0:2) arc(0:120:2) -- (120:1) arc(120:0:1);
        \filldraw[fill=n00] (0:2) -- (0:3) arc(0:40:3) -- (40:2) arc(40:0:2);
        \filldraw[fill=n01] (40:2) -- (40:3) arc(40:80:3) -- (80:2) arc(80:40:2);
        \filldraw[fill=n02] (80:2) -- (80:3) arc(80:120:3) -- (120:2) arc(120:80:2);
        \filldraw[fill=n020] (80:3) -- (80:4) arc(80:120:4) -- (120:3) arc(120:80:3);
        \filldraw[fill=n0200] (80:4) -- (80:5) arc(80:100:5) -- (100:4) arc(100:80:4);
        \filldraw[fill=n0201] (100:4) -- (100:5) arc(100:120:5) -- (120:4) arc(120:100:4);
        \filldraw[fill=n1] (120:1) -- (120:2) arc(120:240:2) -- (240:1) arc(240:120:1);
        \filldraw[fill=n10] (120:2) -- (120:3) arc(120:180:3) -- (180:2) arc(180:120:2);
        \filldraw[fill=n100] (120:3) -- (120:4) arc(120:180:4) -- (180:3) arc(180:120:3);
        \filldraw[fill=n11] (180:2) -- (180:3) arc(180:240:3) -- (240:2) arc(240:180:2);
        \filldraw[fill=n110] (180:3) -- (180:4) arc(180:240:4) -- (240:3) arc(240:180:3);
        \filldraw[fill=n1100] (180:4) -- (180:5) arc(180:210:5) -- (210:4) arc(210:180:4);
        \filldraw[fill=n1101] (210:4) -- (210:5) arc(210:240:5) -- (240:4) arc(240:210:4);
        \filldraw[fill=n2] (240:1) -- (240:2) arc(240:360:2) -- (360:1) arc(360:240:1);

        \node at (-3.5em,3.5em) {\textbf{(b)}};
    }
    \end{tikzpicture}
\end{subfigure}
\begin{subfigure}{0.19\textwidth}
    \centering
    \begin{tikzpicture}[x=0.8em,y=0.8em]
        {
        \definecolor{n0}{HTML}{c1b694}
        \definecolor{n00}{HTML}{d39889}
        \definecolor{n01}{HTML}{c1a178}
        \definecolor{n02}{HTML}{97af7d}
        \definecolor{n020}{HTML}{82a160}
        \definecolor{n0200}{HTML}{7e903b}
        \definecolor{n0201}{HTML}{57974f}
        \definecolor{n1}{HTML}{9abcd6}
        \definecolor{n10}{HTML}{63b4c2}
        \definecolor{n100}{HTML}{24a8bb}
        \definecolor{n11}{HTML}{8aaad8}
        \definecolor{n110}{HTML}{6d9cd8}
        \definecolor{n1100}{HTML}{1b92d3}
        \definecolor{n1101}{HTML}{6888d7}
        \definecolor{n2}{HTML}{d2acc4}
        
        \filldraw[fill=n0] (0:1) -- (0:2) arc(0:120:2) -- (120:1) arc(120:0:1);
        \filldraw[fill=n00] (0:2) -- (0:3) arc(0:40:3) -- (40:2) arc(40:0:2);
        \filldraw[fill=n01] (40:2) -- (40:3) arc(40:80:3) -- (80:2) arc(80:40:2);
        \filldraw[fill=n02] (80:2) -- (80:3) arc(80:120:3) -- (120:2) arc(120:80:2);
        \filldraw[fill=n020] (80:3) -- (80:4) arc(80:120:4) -- (120:3) arc(120:80:3);
        \filldraw[fill=n0200] (80:4) -- (80:5) arc(80:100:5) -- (100:4) arc(100:80:4);
        \filldraw[fill=n0201] (100:4) -- (100:5) arc(100:120:5) -- (120:4) arc(120:100:4);
        \filldraw[fill=n1] (120:1) -- (120:2) arc(120:240:2) -- (240:1) arc(240:120:1);
        \filldraw[fill=n10] (120:2) -- (120:3) arc(120:180:3) -- (180:2) arc(180:120:2);
        \filldraw[fill=n100] (120:3) -- (120:4) arc(120:180:4) -- (180:3) arc(180:120:3);
        \filldraw[fill=n11] (180:2) -- (180:3) arc(180:240:3) -- (240:2) arc(240:180:2);
        \filldraw[fill=n110] (180:3) -- (180:4) arc(180:240:4) -- (240:3) arc(240:180:3);
        \filldraw[fill=n1100] (180:4) -- (180:5) arc(180:210:5) -- (210:4) arc(210:180:4);
        \filldraw[fill=n1101] (210:4) -- (210:5) arc(210:240:5) -- (240:4) arc(240:210:4);
        \filldraw[fill=n2] (240:1) -- (240:2) arc(240:360:2) -- (360:1) arc(360:240:1);

        \node at (-3.5em,3.5em) {\textbf{(c)}};
    }
    \end{tikzpicture}
\end{subfigure}
\begin{subfigure}{0.19\textwidth}
    \centering
    \begin{tikzpicture}[x=0.8em,y=0.8em]
        {
        \definecolor{n0}{HTML}{d0b099}
        \definecolor{n00}{HTML}{cd6b64}
        \definecolor{n01}{HTML}{ba7945}
        \definecolor{n02}{HTML}{b5a675}
        \definecolor{n020}{HTML}{a89756}
        \definecolor{n0200}{HTML}{a18537}
        \definecolor{n0201}{HTML}{938a36}
        \definecolor{n1}{HTML}{8fc1b6}
        \definecolor{n10}{HTML}{63ac85}
        \definecolor{n100}{HTML}{2b9b69}
        \definecolor{n11}{HTML}{62b5b9}
        \definecolor{n110}{HTML}{23a9af}
        \definecolor{n1100}{HTML}{009d9a}
        \definecolor{n1101}{HTML}{009cb0}
        \definecolor{n2}{HTML}{947dcd}
        
        \filldraw[fill=n0] (0:1) -- (0:2) arc(0:120:2) -- (120:1) arc(120:0:1);
        \filldraw[fill=n00] (0:2) -- (0:3) arc(0:40:3) -- (40:2) arc(40:0:2);
        \filldraw[fill=n01] (40:2) -- (40:3) arc(40:80:3) -- (80:2) arc(80:40:2);
        \filldraw[fill=n02] (80:2) -- (80:3) arc(80:120:3) -- (120:2) arc(120:80:2);
        \filldraw[fill=n020] (80:3) -- (80:4) arc(80:120:4) -- (120:3) arc(120:80:3);
        \filldraw[fill=n0200] (80:4) -- (80:5) arc(80:100:5) -- (100:4) arc(100:80:4);
        \filldraw[fill=n0201] (100:4) -- (100:5) arc(100:120:5) -- (120:4) arc(120:100:4);
        \filldraw[fill=n1] (120:1) -- (120:2) arc(120:240:2) -- (240:1) arc(240:120:1);
        \filldraw[fill=n10] (120:2) -- (120:3) arc(120:180:3) -- (180:2) arc(180:120:2);
        \filldraw[fill=n100] (120:3) -- (120:4) arc(120:180:4) -- (180:3) arc(180:120:3);
        \filldraw[fill=n11] (180:2) -- (180:3) arc(180:240:3) -- (240:2) arc(240:180:2);
        \filldraw[fill=n110] (180:3) -- (180:4) arc(180:240:4) -- (240:3) arc(240:180:3);
        \filldraw[fill=n1100] (180:4) -- (180:5) arc(180:210:5) -- (210:4) arc(210:180:4);
        \filldraw[fill=n1101] (210:4) -- (210:5) arc(210:240:5) -- (240:4) arc(240:210:4);
        \filldraw[fill=n2] (240:1) -- (240:2) arc(240:360:2) -- (360:1) arc(360:240:1);

        \node at (-3.5em,3.5em) {\textbf{(d)}};
    }
    \end{tikzpicture}
\end{subfigure}
\begin{subfigure}{0.19\textwidth}
    \centering
    \begin{tikzpicture}[x=0.8em,y=0.8em]
        {
        \definecolor{n0}{HTML}{c1b694}
        \definecolor{n00}{HTML}{c96f59}
        \definecolor{n01}{HTML}{ad7f3c}
        \definecolor{n02}{HTML}{97af7d}
        \definecolor{n020}{HTML}{82a160}
        \definecolor{n0200}{HTML}{7e903b}
        \definecolor{n0201}{HTML}{57974f}
        \definecolor{n1}{HTML}{9abcd6}
        \definecolor{n10}{HTML}{3facbd}
        \definecolor{n100}{HTML}{009bb3}
        \definecolor{n11}{HTML}{8aaad8}
        \definecolor{n110}{HTML}{6d9cd8}
        \definecolor{n1100}{HTML}{1b92d3}
        \definecolor{n1101}{HTML}{6888d7}
        \definecolor{n2}{HTML}{c26ba8}
        
        \filldraw[fill=n0] (0:1) -- (0:2) arc(0:120:2) -- (120:1) arc(120:0:1);
        \filldraw[fill=n00] (0:2) -- (0:3) arc(0:40:3) -- (40:2) arc(40:0:2);
        \filldraw[fill=n01] (40:2) -- (40:3) arc(40:80:3) -- (80:2) arc(80:40:2);
        \filldraw[fill=n02] (80:2) -- (80:3) arc(80:120:3) -- (120:2) arc(120:80:2);
        \filldraw[fill=n020] (80:3) -- (80:4) arc(80:120:4) -- (120:3) arc(120:80:3);
        \filldraw[fill=n0200] (80:4) -- (80:5) arc(80:100:5) -- (100:4) arc(100:80:4);
        \filldraw[fill=n0201] (100:4) -- (100:5) arc(100:120:5) -- (120:4) arc(120:100:4);
        \filldraw[fill=n1] (120:1) -- (120:2) arc(120:240:2) -- (240:1) arc(240:120:1);
        \filldraw[fill=n10] (120:2) -- (120:3) arc(120:180:3) -- (180:2) arc(180:120:2);
        \filldraw[fill=n100] (120:3) -- (120:4) arc(120:180:4) -- (180:3) arc(180:120:3);
        \filldraw[fill=n11] (180:2) -- (180:3) arc(180:240:3) -- (240:2) arc(240:180:2);
        \filldraw[fill=n110] (180:3) -- (180:4) arc(180:240:4) -- (240:3) arc(240:180:3);
        \filldraw[fill=n1100] (180:4) -- (180:5) arc(180:210:5) -- (210:4) arc(210:180:4);
        \filldraw[fill=n1101] (210:4) -- (210:5) arc(210:240:5) -- (240:4) arc(240:210:4);
        \filldraw[fill=n2] (240:1) -- (240:2) arc(240:360:2) -- (360:1) arc(360:240:1);

        \node at (-3.5em,3.5em) {\textbf{(e)}};
    }
    \end{tikzpicture}
\end{subfigure}
\caption{
Comparison of various configurations of the Tree Colors algorithm.
\textbf{(a)} shows the original configuration.
\textbf{(b)} uses adjusted ranges for chroma and luminance to achieve maximum chroma at the leaves without leaving the gamut.
\textbf{(c)} uses a proportional hue split for better discriminative power within sub-trees but sacrificing discriminative power between sub-trees.
\textbf{(d)} uses local interpolation of chroma and luminance to achieve equal visual importance of leaf nodes while sacrificing the equal visual importance of nodes on the same hierarchy level.
\textbf{(e)} combines the proportional hue split with local interpolation. 
}
\label{fig:teaser}
}

\abstract{
To improve the perception of hierarchical structures in data sets, several color map generation algorithms have been proposed to take this structure into account.
But the design of hierarchical color maps elicits different requirements to those of color maps for tabular data.
Within this paper, we make an initial effort to put design rules from the color map literature into the context of hierarchical color maps.
We investigate the impact of several design decisions and provide recommendations for various analysis scenarios.
Thus, we lay the foundation for objective quality criteria to evaluate hierarchical color maps.
} 


\begin{document}

\maketitle

\section{Introduction}
\label{sec:intro}

The investigation of hierarchical data is a frequent subject in many applications, such as analysis of file systems~\cite{borkinEvaluationFilesystemProvenance2013}, biological data, or political maps~\cite{waldinCuttlefishColorMapping2019}.
To improve the perception of the underlying hierarchical structure, several color map generators have been presented in the past to capture the structural properties of such data.
However, we do not have objective criteria to evaluate hierarchical color maps.
The development of color maps in visualization research follows established design rules and many efforts have been made to define quantitative criteria that measure the quality of a color map~\cite{bujackGoodBadUgly2018}, but the inherent structure of hierarchical data elicits different requirements.
Thus, these design rules and measures can not be applied to hierarchical color maps as-is.
Within this paper, we make an initial effort to translate common design rules from the color map literature to the context of hierarchical color maps.
We discuss these quality considerations based on the example of the Tree Colors~\cite{tennekesTreeColorsColor2014} method and present possible adjustments to the algorithm to improve the quality of the resulting color map under certain conditions.

We further consider the focus of the analysis in hierarchical data sets.
To that end, we distinguish between top-down and bottom-up analysis within our elaborations.
We define top-down analysis tasks as those tasks that focus on the top levels of a hierarchy.
Within top-down analysis the hierarchy usually represents nested sets of data points.
Analysts compare these sets and drill down into sets of interest to gain insight into relationships between and within those sets.
Bottom-up analysis, on the other hand, focuses on gaining insight about the hierarchy's leaves.
The hierarchical structure itself is used as context information, or as aggregation mechanism to reduce visual clutter.
Bottom-up analysis is frequently encountered in non-hierarchical visualizations of hierarchical data, where the hierarchy is used to filter the data before making comparisons on the element level.
This distinction in analysis focus elicits different design goals for color maps striving to support either of the two cases.
In summary, our main contributions are:

\begin{itemize}
    \item the translation of color map design rules to the context of hierarchical color maps.
    \item the distinction between top-down and bottom-up analysis of hierarchical data along with its impact on the individual design rules.
    \item the proposition of adjustments to the Tree Colors algorithm to improve the resulting color map's quality under certain conditions.
\end{itemize}

\section{The HCL Color Space}
\label{sec:color}

Within this paper, we elaborate potential design decisions in terms of the HCL color space.
The HCL color space belongs to the class of perceptual color spaces~\cite{zhouSurveyColormapsVisualization2016}.
It is based on the CIELab color space, which was designed to model colors close to the way they are perceived by the human visual system.
HCL is a cylindrical transformation of CIELab with more intuitive dimensions.
This cylinder spans luminance values along its height, chroma along its radius, and hues along its circumference.
Although it should be noted that different hues reach their maximum chroma at different luminances, which causes the color space to have a very irregular shape.
The volume of displayable colors (the gamut) can also differ between display devices.
Nevertheless, this color space is frequently utilized in color map research, because it is almost perceptually uniform~\cite{healeyChoosingEffectiveColours1996} and intuitive to use.
The HCL color space is implemented in popular software libraries such as the \texttt{colorspace}~\cite{zeileisColorspaceToolboxManipulating2020} package for \emph{R} and \emph{Python} as well as \emph{D3.js}~\cite{bostockDataDrivenDocuments2011} and can be interactively explored with hclwizard\footnote{\url{https://hclwizard.org/}}. 

\section{Related Work}

\subsection{Color Map Quality}
\label{sec:quality}

Over the years, the majority of the research effort in design rules for color maps has focused on one dimensional quantitative color maps.
Bujack et al.~\cite{bujackGoodBadUgly2018} present a survey over a large volume of these studies and compile a list of common design rules.
They also map the most prevalent design rules to quantitative quality measures.
Similarly, Bernard et al.~\cite{bernardSurveyTaskbasedQuality2015} present design rules and quantitative measures for two dimensional quantitative color maps.
In contrast, qualitative color maps have received less attention.
The basic design rules have remained the same over the years: keep chroma and luminance constant and vary hue across categories~\cite{harrowerColorBrewerOrgOnline2003, zeileisEscapingRGBlandSelecting2009}.
Based on this approach, many color map generators have been proposed to include additional criteria, such as mark type~\cite{szafirModelingColorDifference2018}, color concept associations~\cite{rathoreEstimatingColorConceptAssociations2020}, color names~\cite{luPalettailorDiscriminableColorization2021}, object proximity in the visualization~\cite{liColorAssignmentOptimization2023}, and aesthetic criteria~\cite{gramazioColorgoricalCreatingDiscriminable2017}.
But so far, such design rules and quality measures have not been applied to hierarchical color maps.
Within this paper, we consider the collection of design rules shared between Bujack et al.~\cite{bujackGoodBadUgly2018} and Bernard et al.~\cite{bernardSurveyTaskbasedQuality2015}, namely \emph{Order}, \emph{Discriminative Power}, \emph{Uniformity}, \emph{Equal Visual Importance}, \emph{Background Sensitivity}, and \emph{Device Independence}.

\subsection{Hierarchical Color Maps}
\label{sec:hierarchical_color_schemes}

The first instance of an inherently hierarchical color map that we could find is the proximity-based coloring introduced by Fua et al.~\cite{fuaHierarchicalParallelCoordinates1999}.
They implemented a multi-scale parallel coordinates visualization that uses a hierarchical color map to color the individual polylines based on their similarity.
To achieve this coloring, they collected all visible elements in a cluster hierarchy using an in-order tree traversal.
Then they linearly assigned hue values to the nodes in the resulting list, causing similar colors to be assigned to a parent and its children.
However, this algorithm is only applicable to binary trees und relies purely on the hue as discriminator.
Nevertheless, most algorithms for hierarchical color maps work similarly.

The structure-based coloring in InterRing~\cite{yangInterRingInteractiveTool2002} applies a bottom-up approach to assign hue values to the leaf nodes first.
The inner nodes of the hierarchy are then assigned the weighted average color of their children.
However, this approach can cause inner nodes of the hierarchy to have the same color as one of their descendants.

The Hyperbolic Wheel~\cite{lamHyperbolicWheelNovel2012} computes the hues of the first hierarchy level based on the relative sub-tree sizes.
Colors are then assigned recursively to children by adding a fixed offset to the parent's hue value.
The brightness of hues is decreased linearly with the depth in the tree, while the saturation is kept constant.
Due to the constant offset, this approach sometimes assigns the same color to nodes on adjacent sub-trees.

The Tree Colors algorithm~\cite{tennekesTreeColorsColor2014} also assigns hues recursively.
It divides the available range of hues among the children of a node, leaving a gap between the nodes to improve the distinguishability of neighboring sub-trees.
Each node is then assigned the center hue value of its range.
Chroma and luminance are varied linearly with the depth of the hierarchy.
While this approach presents a simple solution to the shortcomings of its predecessors, they only consider hierarchies with a depth of three in their publication.

To deal with cases of larger hierarchies, Waldin et al.~\cite{waldinCuttlefishColorMapping2019} proposed Cuttlefish.
Cuttlefish implements a dynamic approach that supports a semantic zoom into the data.
This algorithm is initialized with a set of currently visible groups of items, extracted from the hierarchical data structure.
The range of hues is then divided among the groups and items are placed equidistantly within the hue range of their group.
When the user navigates the view, the coloring is recomputed with the updated set of items.
To improve visual consistency, Cuttlefish rotates the hue ranges to minimize distances between child and parent hues.
This approach improves distinguishability of the visible items significantly, as it can utilize the free color space of invisible sub-trees.
However, because of this, the algorithm can only compute colors for the currently visible set of hierarchy nodes.
Coloring an entire hierarchy with this approach is not possible.

\section{Quality Considerations}

Before we start dealing with the design rules, we will first put the approaches into the context of best practices from the color map literature.
We compare the design goals of hierarchical color maps with those from one dimensional and two dimensional color maps to judge whether the presented approaches apply the available color channels in a manner that is consistent with best practices.
The Tree Colors method~\cite{tennekesTreeColorsColor2014} will serve as example for these considerations.
Tennekes and de Jonge state three design goals for the algorithm:
\begin{enumerate}
     \item Assign unique colors to all nodes within the hierarchy.
     \item Assign similar colors to parent nodes and their children.
     \item Encode the depth of a node in the tree within the node's color.
\end{enumerate}

The first goal matches with design practices for quantitative or categorical color maps.
We do not design color maps that assign the same color to different values or categories either, as this would create ambiguities in lookup tasks.
Yet multiple of the described hierarchical color maps in \Cref{sec:hierarchical_color_schemes} fail to achieve this goal.

The second goal is mirrored by the other approaches discussed in \Cref{sec:hierarchical_color_schemes}, while the third goal is also achieved by the Hyperbolic Wheel~\cite{lamHyperbolicWheelNovel2012} and Cuttlefish~\cite{waldinCuttlefishColorMapping2019}.
These latter two goals can be interpreted as a hierarchical equivalent to the principle of rows and columns introduced by Trumbo~\cite{trumboTheoryColoringBivariate1981}, which states that a two dimensional color map representing a two dimensional data set should allow users to perceive the two data dimensions distinctly.
In that sense, the two dimensions in a hierarchical structure are the vertical dimension along the hierarchy's depth and the horizontal dimension along the nodes on an individual hierarchy level.
The vertical dimension is discrete and quantitative while the horizontal dimension is categorical with groups of items (sibling nodes) that are related.

Best practices from one dimensional color maps imply to utilize luminance or saturation for the vertical dimension with luminance achieving a better discriminative power~\cite{wareRainbowColormapsAre2023}.
For categorical color maps, as mentioned in \Cref{sec:quality}, the best practice is to vary the hue while keeping the other two color attributes constant.
For the horizontal dimension in hierarchical data, this makes even more sense, considering the fact that using hue as primary discriminator introduces hue banding~\cite{wareRainbowColormapsAre2023}.
Hue banding is a perceptual phenomenon that causes a continuous color map to be perceived as several cohesive segments.
As described above, there are groups of related items (siblings) within the horizontal dimension of hierarchical data.
Aligning the hue bands with the structure of the hierarchy, therefore, should improve the perception of these cohesive groups.
These considerations are in concordance with Brewer's recommendations for two dimensional color maps with a quantitative and a categorical dimension~\cite{brewerGuidelinesUsePerceptual1994}.
They also reaffirm the intentions behind the approaches detailed in \Cref{sec:hierarchical_color_schemes}.

However, to find the best implementations for these design goals, we need design rules and quality criteria for hierarchical color maps.
In the following, we investigate the vertical and horizontal dimension of hierarchical color maps in both top-down and bottom-up analysis scenarios with respect to the design rules:
\begin{itemize}
     \item \emph{Order}
     \item \emph{Discriminative Power}
     \item \emph{Uniformity}
     \item \emph{Equal Visual Importance}
     \item \emph{Background Sensitivity}
     \item \emph{Device Independence}
\end{itemize}

\subsection{Order}

The order design rule states that colors in a quantitative color map should imply an ordering.
Thus, it should be possible for viewers to sort a sample of colors based on their perceived order.
For categorical color maps, the best practice states to avoid implying an order~\cite{harrowerColorBrewerOrgOnline2003}.
Thus, to achieve the best color map in terms of order, we need intuitively orderable colors along the vertical dimension, while avoiding such colors along the horizontal dimension.
Research shows that a sampling of hues is difficult to put into an order~\cite{borlandRainbowColorMap2007}, while chroma and luminance can achieve more intuitive orderings.
For this reason, the approach of utilizing hue for the horizontal dimension and luminance and chroma for the vertical dimension can achieve good characteristics in terms of order.
Linearly interpolating luminance and chroma along the depth also makes sure that the color order remains consistent.
Tree Colors further avoids a horizontal order by permuting the colors among siblings, thereby making the perception of order between them more difficult.

\subsection{Discriminative Power}

The discriminative power design rule for quantitative color maps states that the perceptual distance between the colors of the color map should be as large as possible.
The best possible discriminative power would be achieved by a color map that samples the entire color space with an equidistant grid.
But this color map would achieve very poor quality in terms of the other design rules.
Hence, the discriminative power always has to be balanced with the other rules to find the best trade-off.
For the vertical dimension, achieving a good discriminative power seems straightforward.
Because we linearly assign chroma and luminance to the vertical dimension, the discriminative power is only affected by the range of admissible values.
However, these intervals must be adjusted with care, because they also influence other criteria.

Because the sample points along the horizontal dimension lie on a circle in the color space, we improve the discriminative power by increasing the radius of that circle.
This radius is the chroma value.
But the maximum possible chroma differs between luminance values, because of the irregular shape of the gamut in HCL.
Thus, our maximum horizontal discriminative power again depends on the value range of chroma and luminance.
Recent approaches in categorical color map generation improve the discriminative power further by making sure that the sampled hues correspond to different color names and by allowing variations along the other color attributes~\cite{luPalettailorDiscriminableColorization2021}.
However, in the hierarchical context, the latter conflicts with the rows and columns principle and, thus, needs to employed with care.
Another factor to consider is the sampling direction of chroma and luminance.
Because higher values of chroma result in a higher discriminative power, and deeper hierarchy levels usually contain more nodes than those levels closer to the root, it makes intuitive sense to increase the chroma with the depth.
This is especially important when designing for bottom-up analysis, because we need to achieve maximum discriminative power among the hierarchy leaves in this case.
For top-down analysis, the inverse direction may provide better results, because it improves discriminative power among the upper levels of the hierarchy.

\begin{figure}
    \centering
    \begin{tikzpicture}
        \node at (0em,0em) {\includegraphics[width=6em]{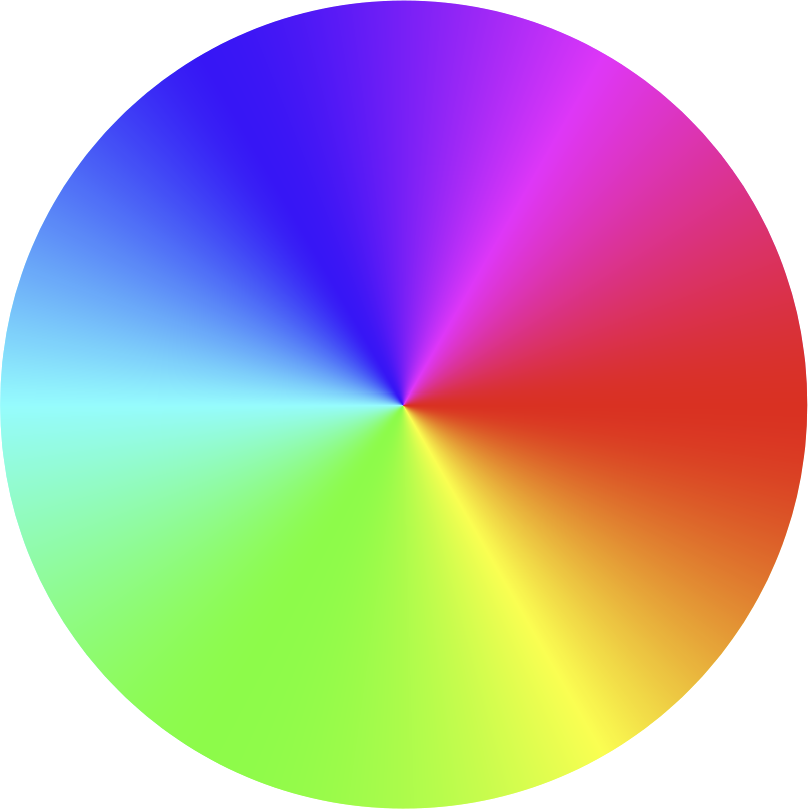}};
        \filldraw[fill=white] circle[radius=1em];
        \draw (0:1em) -- (0:2em) arc(0:120:2em) -- (120:1em);
        \draw (0:2em) -- (0:3em) arc(0:120:3em) -- (120:2em);
        \draw (120:1em) -- (120:2em) arc(120:240:2em) -- (240:1em);
        \draw (120:2em) -- (120:3em) arc(120:180:3em) -- (180:2em);
        \draw (180:2em) -- (180:3em) arc(180:240:3em) -- (240:2em);
        \draw (240:1em) -- (240:2em) arc(240:360:2em) -- (360:1em);
        \draw (240:2em) -- (240:3em) arc(240:280:3em) -- (280:2em);
        \draw (280:2em) -- (280:3em) arc(280:320:3em) -- (320:2em);
        \draw (320:2em) -- (320:3em) arc(320:360:3em) -- (360:2em);

        \node at (12em,0em) {\includegraphics[width=6em]{ColorWheel.png}};
        \filldraw[fill=white] (12em,0em) circle[radius=1em];
        \draw ($(0:1em)+(12em,0em)$) -- ($(0:2em)+(12em,0em)$) arc(0:60:2em) -- ($(60:1em)+(12em,0em)$);
        \draw ($(0:2em)+(12em,0em)$) -- ($(0:3em)+(12em,0em)$) arc(0:60:3em) -- ($(60:2em)+(12em,0em)$);
        \draw ($(60:1em)+(12em,0em)$) -- ($(60:2em)+(12em,0em)$) arc(60:180:2em) -- ($(180:1em)+(12em,0em)$);
        \draw ($(60:2em)+(12em,0em)$) -- ($(60:3em)+(12em,0em)$) arc(60:120:3em) -- ($(120:2em)+(12em,0em)$);
        \draw ($(120:2em)+(12em,0em)$) -- ($(120:3em)+(12em,0em)$) arc(120:180:3em) -- ($(180:2em)+(12em,0em)$);
        \draw ($(180:1em)+(12em,0em)$) -- ($(180:2em)+(12em,0em)$) arc(180:360:2em) -- ($(360:1em)+(12em,0em)$);
        \draw ($(180:2em)+(12em,0em)$) -- ($(180:3em)+(12em,0em)$) arc(180:240:3em) -- ($(240:2em)+(12em,0em)$);
        \draw ($(240:2em)+(12em,0em)$) -- ($(240:3em)+(12em,0em)$) arc(240:300:3em) -- ($(300:2em)+(12em,0em)$);
        \draw ($(300:2em)+(12em,0em)$) -- ($(300:3em)+(12em,0em)$) arc(300:360:3em) -- ($(360:2em)+(12em,0em)$);

        \node at (-4em,3.5em) {\textbf{(a)}};
        \node at (8em,3.5em) {\textbf{(b)}};
    \end{tikzpicture}
    \caption{
    Comparison of even \textbf{(a)} and proportional \textbf{(b)} hue split variants in imbalanced hierarchies.
    Sizes of arc segments indicate the range of hues assigned to each node.
    }
    \label{fig:split_mode}
\end{figure}

We also need to consider whether our design requires maximum discriminative power between groups of siblings or within groups of siblings.
The Tree Colors method, for example, leaves gaps in the assigned hue ranges to improve discriminative power among separate sub-trees.
But large gaps reduce the fraction of hue that is available to discriminate between the colors within each sub-tree.
Thus, designers must tune this hue fraction to an appropriate trade-off.
The algorithm also splits the available hue evenly between all siblings, irrespective of their sub-trees' sizes.
This improves the discriminative power among the siblings' sub-trees, but reduces the discriminative power within the larger sub-trees.
For this reason, alternative approaches split the hue range proportionally to the sub-tree size~\cite{fuaHierarchicalParallelCoordinates1999, lamHyperbolicWheelNovel2012, waldinCuttlefishColorMapping2019, yangInterRingInteractiveTool2002}.
\Cref{fig:split_mode} shows a comparison of the two hue split variants.
Notice how in the even case, a sub-tree with a single leaf is assigned the same amount of hue as the larger sub-tree with three leaves. 
For visualizations supporting bottom-up analysis, the proportional split is generally preferable, because the identification of individual elements is most important, while the hierarchical structure of the data set serves merely as context information.
The opposite is true for top-down analysis scenarios.

Dynamic approaches, such as Cuttlefish~\cite{waldinCuttlefishColorMapping2019}, also improve the discriminative power of the visible elements by re-assigning colors upon view changes.
However, this reduces the discriminative power between the currently visible elements and those that were visible in the previous view, leading in the extreme case to different elements with exactly matching colors over separate views.

\subsection{Uniformity}

The uniformity design rule states that perceptual color distances should represent the distances between data values.
For quantitative color maps, the distance between two colors should ideally be proportional to the distance between the values they represent.
For the vertical dimension in a hierarchical color map, this implies that the discrete depth values should be represented by equidistant colors, which we achieve by interpolating linearly.
For qualitative color maps, the chosen colors should ideally be equidistant.
However, this conflicts with the approach of assigning hues of different color names to maximize discriminative power.
Furthermore, we need to consider the color distance within groups of siblings and across separate sub-trees.
Within groups of siblings we can apply equidistant colors.
The introduction of a gap in hue between neighboring sub-trees achieves a larger distance between non-sibling nodes than between siblings.
This improves uniformity, because these nodes are conceptually further apart.
But this conceptual distance is difficult to quantify.
For this reason it is also difficult to define uniformity across separate sub-trees.
Tree Colors resembles a concept similar to that of a general tree distance, in which the distance between two nodes is given by the depth from their closest common ancestor.

\subsection{Equal Visual Importance}

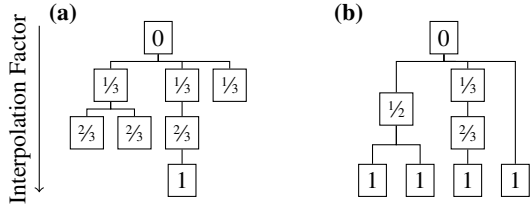
\begin{figure}
    \centering
    \begin{tikzpicture}
        \node[draw,name=groot] at (0em,0em) {0};
        \node[draw,name=g0] at (-2em,-2em) {\sfrac{1}{3}};
        \node[draw,name=g00] at (-3em,-4em) {\sfrac{2}{3}};
        \node[draw,name=g01] at (-1em,-4em) {\sfrac{2}{3}};
        \node[draw,name=g1] at (1em,-2em) {\sfrac{1}{3}};
        \node[draw,name=g10] at (1em,-4em) {\sfrac{2}{3}};
        \node[draw,name=g100] at (1em,-6em) {1};
        \node[draw,name=g2] at (3em,-2em) {\sfrac{1}{3}};
        \draw (groot.south) |- ($(groot.south)!0.5!(g0.north)$) -| (g0.north);
        \draw (groot.south) |- ($(groot.south)!0.5!(g1.north)$) -| (g1.north);
        \draw (groot.south) |- ($(groot.south)!0.5!(g2.north)$) -| (g2.north);
        \draw (g0.south) |- ($(g0.south)!0.5!(g00.north)$) -| (g00.north);
        \draw (g0.south) |- ($(g0.south)!0.5!(g01.north)$) -| (g01.north);
        \draw (g1) -- (g10);
        \draw (g10) -- (g100);

        \node[draw,name=lroot] at (12em,0em) {0};
        \node[draw,name=l0] at (10em,-3em) {\sfrac{1}{2}};
        \node[draw,name=l00] at (9em,-6em) {1};
        \node[draw,name=l01] at (11em,-6em) {1};
        \node[draw,name=l1] at (13em,-2em) {\sfrac{1}{3}};
        \node[draw,name=l10] at (13em,-4em) {\sfrac{2}{3}};
        \node[draw,name=l100] at (13em,-6em) {1};
        \node[draw,name=l2] at (15em,-6em) {1};
        \draw (lroot.south) |- ($(lroot.south)!0.5!(l1.north)$) -| (l0.north);
        \draw (lroot.south) |- ($(lroot.south)!0.5!(l1.north)$) -| (l1.north);
        \draw (lroot.south) |- ($(lroot.south)!0.5!(l1.north)$) -| (l2.north);
        \draw (l0.south) |- ($(l0.south)!0.5!(l00.north)$) -| (l00.north);
        \draw (l0.south) |- ($(l0.south)!0.5!(l01.north)$) -| (l01.north);
        \draw (l1) -- (l10);
        \draw (l10) -- (l100);

        \node at (-4em,1em) {\textbf{(a)}};
        \node at (8em,1em) {\textbf{(b)}};
        \draw[->] (-5em,0.5em) -- (-5em,-6.5em);
        \node[anchor=south,rotate=90] at (-5em,-3em) {Interpolation Factor};
    \end{tikzpicture}
    \caption{
    Comparison of global \textbf{(a)} and local \textbf{(b)} interpolation variants for chroma and luminance in imbalanced hierarchies.
    Interpolation factors of individual nodes are written inside the nodes and encoded by their vertical position.
    }
    \label{fig:interpolation_mode}
\end{figure}

Bernard et al.~\cite{bernardSurveyTaskbasedQuality2015} define equal visual importance as \enquote{the requirement that all colors are equally salient}.
Different color temperatures, hues, chroma and luminance values may draw the attention of viewers to certain elements of the visualization more than to others.
When all elements of the visualization are of equal importance to the analysis, they should also be represented by colors with an equal capability to draw the viewer's attention.
Along the vertical dimension, we need to consider the impact of chroma and luminance variations.
Generally, brighter and more saturated colors draw more attention than darker, muted colors.
Tree Colors assigns chroma and luminance in an inverse relationship with the depth of the hierarchy.
One is always increased while the other is decreased.
Intuitively, this can improve the criterion of equal visual importance, because an increase in one of the two color attributes can be compensated by a decrease in the other.
Although we are not aware of any empirical research that confirms this intuition.
But this approach may conflict with the discriminative power and background sensitivity criteria in certain analysis scenarios.
We may also need to consider the analysis focus.
Tree Colors computes chroma and luminance linearly with the depth in the hierarchy.
If the hierarchy contains leaves at different depths, these leaves are assigned different chroma and luminance values.
This can cause a different perception of importance across the leaves, which is problematic in bottom-up analysis scenarios.
Here, it may be advantageous to assign a fixed value for chroma and luminance to all leaves.
However, because the different branches of the hierarchy are of different lengths, we need to interpolate the chroma and luminance values for each branch locally.
\Cref{fig:interpolation_mode} illustrates the difference between local and global interpolation for imbalanced hierarchies.
With local interpolation, the vertical dimension of the color map is only uniform within an individual branch.
Across separate branches, the design rule of uniformity is violated and the color map no longer satisfies equal visual importance across a single inner hierarchy level.

Along the horizontal dimension, chroma and luminance are kept constant, but certain hues may appear more prominent than others.
To draw viewers' attention, important information is often communicated via signal colors, such as red to imply danger.
The straightforward approach to prevent the representation of data by signal colors is to exclude these colors from the range of admissible hues.
But introducing gaps in the hue range reduces the discriminative power of the remaining color space and conflicts with uniformity.

\subsection{Background Sensitivity}

Background sensitivity is described by the contrast between colored elements of the visualization and the background.
There is also the corresponding design rule of foreground sensitivity, which considers the contrast between the visualization colors and foreground elements such as axes or text labels.
Because background and foreground elements are usually drawn with colors close to white or black, the contrast to these elements is mainly determined by the luminance.
If no foreground elements exist, we can use a darker portion of the luminance range in light application themes and a brighter portion in dark application themes.
If the visualization contains foreground elements that overlap with the hierarchy's colors, we need to use a narrower portion from the center of the luminance axis to achieve good contrast to both ends of the scale.
In this case, it can also help to increase the minimum chroma to improve the contrast.
If foreground elements are drawn exclusively on the colored elements, we can also chose to draw foreground elements in the same color as the background to increase our available range of luminance.
Because the background sensitivity design rule restricts our luminance range, it can conflict with the discriminative power along both dimensions.
For top-down analysis, we need to guarantee a good background sensitivity for the upper hierarchy levels, while we need to guarantee the same for the leaves in bottom-up analysis scenarios.
Thus, the direction of the luminance interpolation should be adapted to the more likely scenario.

\subsection{Device Independence}

Colors sampled outside of a device's gamut are clipped to the gamut's boundary~\cite{morelandDivergingColorMaps2009}.
This can result in different representations of the same specified HCL color on devices with different gamuts, which can deteriorate the color map's quality in terms of equal visual importance, uniformity and discriminative power.
Thus, the device independence design rule states that colors should always be sampled from within a standardized cross-device gamut such as sRGB.
This limits the space of available colors and may negatively impact the discriminative power of the resulting color map, but it guarantees consistent color representation across different devices.

\section{Observations}

During our investigation of the design rules, we discovered configurations of Tree Colors that provide good results.
As starting point for designers, we provide these parameters in \Cref{tab:observations} as well as a summary of our findings as design cheat sheet in the supplementary material.
For small hierarchies, such as the examples considered by Tennekes and de Jonge, the difference in requirements regarding the analysis focus is negligible.
Under those circumstances, the recommended parameters of the original Tree Colors method produce satisfactory results~\cite{tennekesTreeColorsColor2014, tennekesErrataTreeColors2015}.
But designers should keep in mind that this configuration leaves the sRGB gamut and, therefore, achieves poor device independence and equal visual importance.
For larger hierarchies, note that the maximum chroma is larger for the dark theme than the light theme, which causes the dark theme to achieve better discriminative power.
The exclusion of hues should be used sparingly both in the number of excluded hues and the angle of each excluded hue-slice.
If certain hues must be excluded, we recommend to exclude slices of about $12^{\circ}$.
For very large hierarchies, with hundreds of nodes, dynamic approaches are necessary.

\begin{table}
     \small
     \center
     \SetTblrInner{rowsep=0pt}
     \begin{tblr}{%
          colspec={c r | c c | c c},%
          column{1}={bg=fhg, fg=white, font=\bfseries},%
          row{1}={bg=fhg, fg=white, font=\bfseries},%
          cell{3-8}{2}={fhglite},%
          cell{3,6}{3,4,5,6,7}={fhglite}%
     }
          \SetCell[c=2]{c} Hierarchy Size & & \SetCell[c=2]{c} Small & & \SetCell[c=2]{c} Larger & \\
          \SetCell[c=2]{c} Hue Fraction & & \SetCell[c=2]{c} $0.75$ & & \SetCell[c=2]{c} $0.9$ & \\
          \hline \hline
          \SetCell[r=3]{c} {Application \\ Theme} & & Lum. & Chroma & Lum. & Chroma \\
           & Light & \SetCell[c=2]{c} Additive Color & & $[95, 57]$ & $[10, 45]$ \\
           & Dark & \SetCell[c=2]{c} Subtractive Color & & $[26, 76]$ & $[20, 59]$ \\
          \hline \hline
          \SetCell[r=3]{c} {Analysis \\ Focus} & & Interp. & Hue Split & Interp. & Hue Split \\
           & Top-Down & \SetCell[r=2,c=2]{c,m} any & & global & even \\
           & Bottom-Up & & & local & prop. \\
     \end{tblr}
     \caption{%
          Good configurations of Tree Colors.
          Application theme and analysis focus can be independently combined to yield eight different configurations.
          Luminance and chroma are given as intervals starting at the top of the hierarchy.
     }
     \label{tab:observations}
\end{table}

\section{Conclusion}

Within this paper, we have translated the most prevalent design rules from the color map literature into the context of hierarchical color maps.
We have further investigated the impact of application theme and analysis focus on the color map quality in regard to the individual design rules and have provided recommendations for the adaptation of Tree Colors to improve the quality in different scenarios.
We thus lay the foundation for an objective discussion of hierarchical color map quality.
As next step, we must determine how to quantitatively measure the quality according to these design rules and design appropriate benchmark data sets to evaluate color map generation algorithms.
As an extension to this work, there are also many design rules in the literature that we have yet to translate.

\acknowledgments{
This work is funded by the German Federal Ministry for Education and Research (BMBF) and the Hessian Ministry for Science and Art (HMWK) within their joint support of the National Research Center for Applied Cybersecurity ATHENE.}

\bibliographystyle{abbrv-doi-hyperref}

\bibliography{main}
\end{document}


\maketitle

\section{Order}
\label{rule:order}

\subsection*{Horizontal}

\subsubsection*{Goal}
Prevent a perceptible ordering.

\subsubsection*{Optimizations}
\begin{enumerate}
  \item Permute hues.
\end{enumerate}  

\subsection*{Vertical}

\subsubsection*{Goal}
Assign ordered colors.

\subsubsection*{Optimizations}
\begin{enumerate}
  \setcounter{enumi}{1}
  \item Interpolate luminance and chroma linearly.
  \item Use same direction for interpolation of both values.
  \begin{itemize}
    \item Tradeoff with \ref{rule:discriminative_power}.5, \ref{rule:equal_visual_importance}.3, \ref{rule:background_sensitivity}.2, \ref{rule:background_sensitivity}.4.
  \end{itemize}
\end{enumerate}

\section{Discriminative Power}
\label{rule:discriminative_power}

\subsubsection*{Goal}
Increase perceptual distance between sampled colors.

\subsection*{Horizontal}

\subsubsection*{Optimizations}
\begin{enumerate}
  \item Use entire range of hues.
  \begin{itemize}
    \item Tradeoff with \ref{rule:uniformity}.2, \ref{rule:equal_visual_importance}.1.
  \end{itemize}
  \item Increase chroma.
  \begin{itemize}
    \item Tradeoff with \ref{rule:equal_visual_importance}.2, \ref{rule:background_sensitivity}.1, \ref{rule:background_sensitivity}.2, \ref{rule:background_sensitivity}.4, \ref{rule:device_independence}.1.
  \end{itemize}
  \item Sample hues from different color names.
  \begin{itemize}
    \item Tradeoff with \ref{rule:uniformity}.1.
  \end{itemize}
  \item Allow variations of luminance and chroma.
  \begin{itemize}
    \item Tradeoff with rows and columns principle.
  \end{itemize}
  \item Tradeoff between leaf level and upper level.
  \begin{enumerate}
    \item Leaf level
    \begin{itemize}
      \item Set chroma interpolation direction for max value at the leaves.
    \end{itemize}
    \item Upper level
    \begin{itemize}
      \item Set chroma interpolation direction for min value at the leaves.
    \end{itemize}
  \end{enumerate}
  \begin{itemize}
    \item Tradeoff with \ref{rule:order}.3, \ref{rule:equal_visual_importance}.3.
  \end{itemize}
  \item Tradeoff between groups and within groups.
  \begin{enumerate}
    \item Between groups
    \begin{itemize}
      \item Decrease hue fraction.
      \item Use even hue split.
    \end{itemize}
    \item Within groups
    \begin{itemize}
      \item Increase hue fraction.
      \item Use proportional hue split.
    \end{itemize}
  \end{enumerate}
  \begin{itemize}
    \item Tradeoff with \ref{rule:uniformity}.2.
  \end{itemize}
  \item Tradeoff between views or within views (dynamic approach).
  \begin{enumerate}
    \item Between views
    \begin{itemize}
      \item Limit variations in hue re-assignment.
    \end{itemize}
    \item Within views
    \begin{itemize}
      \item Allow greater variations in hue re-assignment.
    \end{itemize}
  \end{enumerate}
\end{enumerate}

\subsection*{Vertical}

\subsubsection*{Optimizations}
\begin{enumerate}
  \setcounter{enumi}{7}
  \item Increase sampling interval of chroma and luminance.
  \begin{itemize}
    \item Tradeoff with \ref{rule:equal_visual_importance}.2, \ref{rule:background_sensitivity}.1, \ref{rule:background_sensitivity}.2, \ref{rule:background_sensitivity}.3, \ref{rule:device_independence}.1.
  \end{itemize}
\end{enumerate}

\section{Uniformity}
\label{rule:uniformity}

\subsection*{Horizontal}

\subsubsection*{Goal}
Assign perceptually equidistant colors within groups with larger distances between groups. For example, conceptually similar to a tree distance.

\subsubsection*{Optimizations}
\begin{enumerate}
  \item Interpolate hues linearly within groups.
  \begin{itemize}
    \item Tradeoff with \ref{rule:discriminative_power}.3, \ref{rule:equal_visual_importance}.1.
  \end{itemize}
  \item Insert gaps in hue between siblings.
  \begin{itemize}
    \item Tradeoff with \ref{rule:discriminative_power}.1, \ref{rule:discriminative_power}.6
  \end{itemize}
\end{enumerate}

\clearpage
\subsection*{Vertical}

\subsubsection*{Goal}
Assign chroma and luminance proportional to the depth.

\subsubsection*{Optimizations}
\begin{enumerate}
  \setcounter{enumi}{2}
  \item Interpolate luminance and chroma linearly.
\end{enumerate}

\section{Equal Visual Importance}
\label{rule:equal_visual_importance}

\subsection*{Horizontal}

\subsubsection*{Goal}
Prevent signal color hues. Assign colors with equal luminance and chroma.

\subsubsection*{Optimizations}
\begin{enumerate}
  \item Exclude signal color hues from sampling.
  \begin{itemize}
    \item Tradeoff with \ref{rule:discriminative_power}.1, \ref{rule:uniformity}.1.
  \end{itemize}
  \item Sample hues from a circle in the color space (constant luminance and chroma).
  \begin{itemize}
    \item Tradeoff with \ref{rule:discriminative_power}.2, \ref{rule:discriminative_power}.8, \ref{rule:device_independence}.1.
  \end{itemize}
\end{enumerate}

\subsection*{Vertical}

\subsubsection*{Goal}
Assign balance the attention steering properties of luminance and chroma throughout the entire hierarchy.

\subsubsection*{Optimizations}
\begin{enumerate}
  \setcounter{enumi}{2}
  \item Assign chroma and luminance in an inverse relationship.
  \begin{itemize}
    \item Tradeoff with \ref{rule:order}.3, \ref{rule:discriminative_power}.5, \ref{rule:background_sensitivity}.4.
  \end{itemize}
  \item Tradeoff between leaf level and same-depth level.
  \begin{enumerate}
    \item Leaf level
    \begin{itemize}
      \item Use local interpolation variant for chroma and luminance.
    \end{itemize}
    \item Same-depth level
    \begin{itemize}
      \item Use global interpolation variant for chroma and luminance.
    \end{itemize}
  \end{enumerate}
\end{enumerate}

\section{Background Sensitivity}
\label{rule:background_sensitivity}

\subsubsection*{Goal}
Assign colors with sufficient contrast to the background (and the foreground).

\subsubsection*{Optimizations}
\begin{enumerate}
  \item Restrict luminance interpolation interval to achieve contrast with background.
  \begin{itemize}
    \item Tradeoff with \ref{rule:order}.3, \ref{rule:discriminative_power}.2, \ref{rule:discriminative_power}.8.
  \end{itemize}
  \item Restrict luminance interpolation interval to achieve contrast with foreground (if necessary).
  \begin{itemize}
    \item Tradeoff with \ref{rule:order}.3, \ref{rule:discriminative_power}.2, \ref{rule:discriminative_power}.8.
  \end{itemize}
  \item Increase minimum chroma.
  \begin{itemize}
    \item Tradeoff with \ref{rule:discriminative_power}.8.
  \end{itemize}
  \item Tradeoff between leaf level and upper level.
  \begin{enumerate}
    \item Leaf level
    \begin{itemize}
      \item Set luminance interpolation direction for max value at the leaves.
    \end{itemize}
    \item Upper level
    \begin{itemize}
      \item Set luminance interpolation direction for min value at the leaves.
    \end{itemize}
  \end{enumerate}
  \begin{itemize}
    \item Tradeoff with \ref{rule:order}.3, \ref{rule:discriminative_power}.2, \ref{rule:equal_visual_importance}.3.
  \end{itemize}
\end{enumerate}

\section{Device Independence}
\label{rule:device_independence}

\subsubsection*{Goal}
Sample colors from within a cross-device gamut (such as sRGB).

\subsubsection*{Optimizations}
\begin{enumerate}
  \item Restrict luminance and chroma interpolation to stay in the gamut.
  \begin{itemize}
    \item Tradeoff with \ref{rule:discriminative_power}.2, \ref{rule:discriminative_power}.8, \ref{rule:equal_visual_importance}.2.
  \end{itemize}
\end{enumerate}

\section{Objectives in Given Scenarios}

\subsection{Top-Down Analysis}

\begin{itemize}
  \item Increase discriminative power in upper hierarchy levels.
  \begin{itemize}
    \item Set chroma interpolation direction for min values at the leaves.
  \end{itemize}
  \item Increase between-group discriminative power.
  \begin{itemize}
    \item Decrease the hue fraction.
    \item Use an even hue split.
  \end{itemize}
  \item Provide equal visual importance on same-depth level.
  \begin{itemize}
    \item Use global interpolation for luminance and chroma.
  \end{itemize}
  \item Increase background sensitivity in upper hierarchy levels.
  \begin{itemize}
    \item Set luminance interpolation direction for max values at the leaves in the light theme.
    \item Set luminance interpolation direction for min values at the leaves in the dark theme.
  \end{itemize}
\end{itemize}

\subsection{Bottom-Up Analysis}

\begin{itemize}
  \item Increase discriminative power on the leaf level.
  \begin{itemize}
    \item Set chroma interpolation direction for max value at the leaves.
  \end{itemize}
  \item Increase within-group discriminative power.
  \begin{itemize}
    \item Increase the hue fraction.
    \item Use a proportional hue split.
  \end{itemize}
  \item Provide equal visual importance on the leaf level.
  \begin{itemize}
    \item Use local interpolation for luminance and chroma.
  \end{itemize}
  \item Increase background sensitivity on the leaf level.
  \begin{itemize}
    \item Set luminance interpolation direction for min values at the leaves in the light theme.
    \item Set luminance interpolation direction for max values at the leaves in the dark theme.
  \end{itemize}
\end{itemize}